\documentclass[%
 preprint,
 amsmath,amssymb,
 aps, prl,superscriptaddress
]{revtex4-1}

\usepackage{dcolumn}
\usepackage{bm}

\usepackage{amssymb}
\usepackage{amsmath} 
\usepackage[subnum]{cases}
\usepackage{empheq}
\usepackage{siunitx}
\usepackage{amsmath}

\usepackage[pdftex]{graphicx}

\usepackage{color}

\begin{document}
\preprint{APS/123-QED}
\title{Optimal switching strategies for stochastic geocentric/egocentric navigation}

\author{O. Peleg}
\affiliation{John A. Paulson School of Engineering and Applied Sciences, Harvard University, Cambridge, Massachusetts 02138,USA}
\author{L. Mahadevan}
 \email{lm@seas.harvard.edu}
 \affiliation{John A. Paulson School of Engineering and Applied Sciences, Harvard University, Cambridge, Massachusetts 02138,USA}
\affiliation{Departments of Physics, and Organismic and Evolutionary Biology, Wyss Institute and Kavli Institute, Harvard University, Cambridge, Massachusetts 02138,USA }
\date{\today}
\begin{abstract}

Animals use a combination of egocentric navigation driven by the internal integration of environmental cues, interspersed with geocentric course correction and reorientation, often with uncertainty in sensory acquisition of information, planning and execution. Inspired directly by observations of dung beetle navigational strategies that show switching between geocentric and egocentric strategies, we consider the question of optimal strategies for the navigation of an agent along a preferred direction in the presence of multiple sources of noise. We address this using a model that takes the form of a correlated random walk at short time scales that is interspersed with reorientation events that yields a biased random walks at long time scales. We identify optimal alternation schemes and characterize their robustness in the context of noisy sensory acquisition, and performance errors linked with variations in environmental conditions and agent-environment interactions. 
\end{abstract}
\pacs{Valid PACS appear here}
\maketitle

Navigation in complex uncertain environments requires information about environmental cues (landmarks) along with the ability to memorize and execute intended plans based on these cues. It is thus often accompanied by several cognitive demanding activities, such as multi-sensory acquisition and integration, locomotion planning, and motor control.  As organisms have finite cognitive and computational resources, they must therefore multitask and develop optimal schemes to dynamically allocate resources to different tasks \cite{Warrant:2010fc,Anonymous:1995uq, Wiltschko:2005fz, Papastamatiou:2011bf}. Similar demands are also placed on their artificial analogs, autonomous vehicles such as robots, self-driving cars, and even deep-space craft \cite{Wood:2008wu,Quadrelli:2015dr}. Typical strategies for navigation involve a combination of egocentric and geocentric schemes. In the first, the organism uses information that it has acquired somehow to move along a path without any realtime feedback. In the second, the organism constantly probes its location relative to environmental cues and adjusts its strategy in real time. The first requires perfect memory and can be efficient, but can succeed only in a constant environment; the second requires continuous course corrections and is often slow, but can cope with fluctuating environments. Organisms switch between these strategies to navigate accurately in a variable environment, and understanding the switching strategies of organisms in their natural environments is a basic problem in neuroethology.
\begin{figure}
   \centering
   \includegraphics[width=10cm]{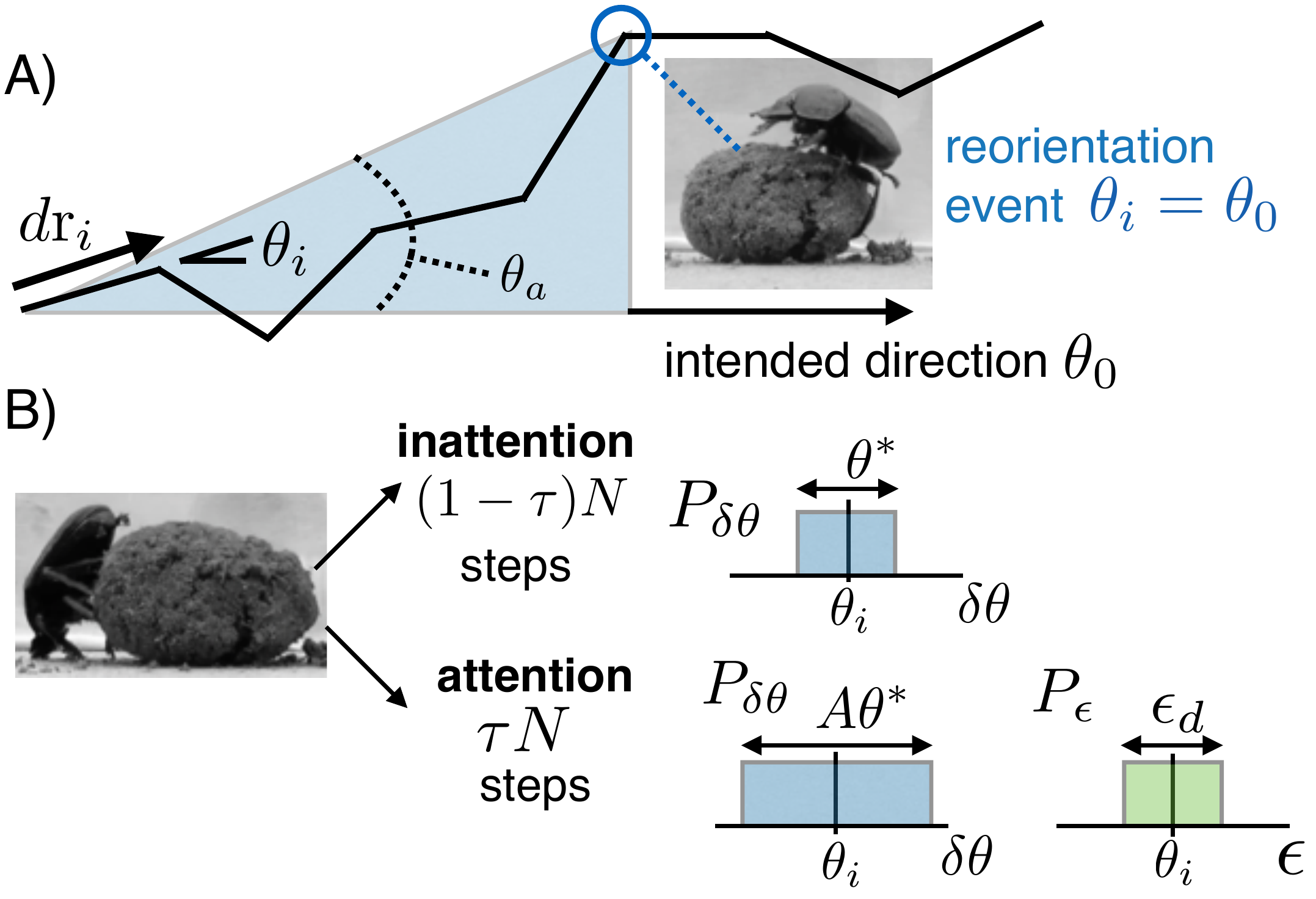} 
   \caption{A) Schematic of the navigation model for the dung beetle. The  beetle {(agent)} walks along a correlated random walk, with errors introduced at every step. When the accumulated error exceeds a threshold $\theta_a$, the {(agent)} resets its orientation and continues. B) Strategies and associated errors. If the {agent} is {inattentive to navigation cues during a step}, the standard deviation of the distribution of angles is $\theta^*$, else it increases by a factor $A>1$ and becomes $A\theta^*$ over the attentive steps $(1-\tau)N$. Similarly, there is an error associated with sensory acquisition that is assumed to be uniformly distributed with a standard deviation $\epsilon_d$. }\vspace{-10pt} 
   \label{fig_model1}
\end{figure}
Dung beetles provide a well studied example of this switching behavior during navigation {\cite{Byrne:2011fu}}. Foraging beetles look for nutrient-rich dung, and then attempt to roll a dung-ball along a straight path radially away from the pile \cite{Warrant:2010fc} {with the aim of providing food for their brood}. Beetles acquire navigation information using a variety of long-range cues before initiating a roll, and then push the dung ball while walking backwards with their hind legs in contact with the dung.  However, they stop intermittently and get atop the ball and walk on it to reorient themselves before continuing to roll the dung-ball. The reorientation behavior of the beetle is visually mediated, triggered both by active and passive deviations off-course and by visual cues \cite{Baird:2012vn} that include the position of the sun, the moon, and associated light polarization patterns. Not having to rely on terrestrial landmarks allows the beetle to travel arbitrarily far away from its starting point without large directional errors \cite{Cheung:2007hna}, thus making the nature of the navigational task fundamentally different from that of homing insects. Furthermore, the switching behavior between runs and reorientations is a function of the environment; the more uncertain the environment as characterized by its physical roughness, the more frequent the reorientations \cite{Baird:2012vn}. Stopping to reorient after each step makes for slow progress, but guarantees a straighter route. Alternatively, not stopping at all avoids the loss of time associated with reorientation, but leads to trajectories that deviate significantly from the intended bearing. This leads to a natural question of the optimal rate of switching between the egocentric strategy and the geocentric one. In this Letter, we introduce a model for the movement of the beetle, or a navigational agent, that is associated with paths that are a characteristic of correlated diffusion on short time scales, and biased diffusion on long time scales \cite{Codling:2008bm}. This allows us to pose and solve an optimization problem for the most efficient switching strategy and characterize its robustness in the presence of noisy sensory acquisition and in rough environments, with relevance to questions that go beyond the original motivation for the problem.

\begin{figure}
   \centering
   \includegraphics[width=11cm]{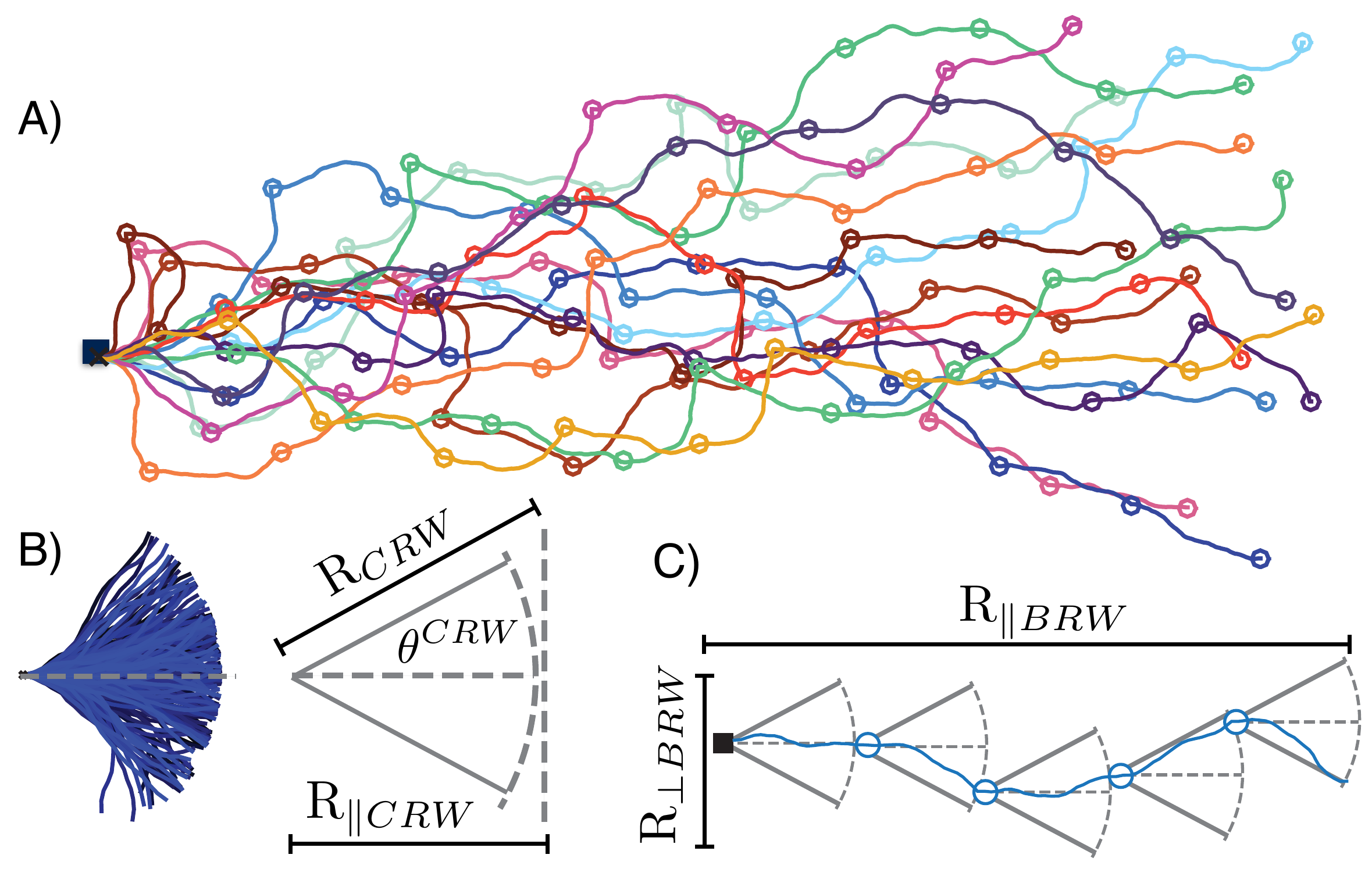} 
   \caption{A) Examples of trajectories obtained via random walk simulations with fixed reorientation interval. Different realizations are shown in different colors, and reorientation points are shown as empty circles. The origin is marked with a black square. $\theta^*=\pi/12$, $N=100$ and $n=20$. B) CRWs of $n=20$ steps and $\theta^*=\pi/12$. Each realization is shown in a different shade of blue. The area accessible to the agent between reorientation events is shown as a circular sector. C) The entire walk is described as a BRW with step size and turning angles defined by the CRW statistics.} \vspace{-10pt} 
   \label{fig_model2}
\end{figure}

Inspired by the beetle, we consider an agent that performs a random walk interspersed by reorientation events, in which its heading direction is reset. We expect that owing to finite cognitive and attention capacity, information gathered while rolling the ball leads to { a bigger} acquisition error relative to that when it is reorienting, { i.e.~the detected orientation is $\theta_i+\epsilon$;  we assume that this dynamic acquisition error, $\epsilon$, is drawn from a uniform distribution of headings $[0,\epsilon_d]$, where $\epsilon\in[0,\pi]$.} Between reorientation events, we assume that the random walk is correlated orientationally, with a mean corresponding to the current orientation and a uniform distribution of angular errors in the range $[0,\theta^*]$, { where $\theta^*\in[0,\pi]$.} The agent can acquire (visual) sensory information in two modes: when it is rolling the ball, or when it reorients. Memory degradation, execution errors and environmental uncertainties translate to larger accumulated errors, i.e.~large $\theta^*$, so that the agent must stop to reorient more often, consistent with observations of beetles \cite{Baird:2012vn}. This suggests that reorientation is triggered by a threshold deviation from the preferred heading, which we denote by $\theta_a$, { where $\theta_a\in[0,\pi]$}. However, reorientation events do not occur at the same angular deviation from the original bearing \cite{Baird:2012vn}, and therefore requires the addition of further triggering mechanisms. Indeed, beetles detect gradual deviation better than abrupt ones \cite{Baird:2012vn}. This suggests that beetles do not pay attention to navigational signals at all times due to finite cognitive resources; when paying attention to navigational cues, motor control suffers, and vice-versa.   To quantify this, we  define the attention span, $\tau$, { where $\tau\in[0,1]$}, as the fraction of time during when the agent pays relatively more attention to navigational cues, with a wider distribution of turning angles drawn from a uniform distribution $[0,A\theta^*]$ (with $A>1$) relative to the fraction of time $1-\tau$, when the distribution of angles is in the range $[0,\theta^*]$. These sources of stochasticity in sensing, attention and movement as the agent interacts with the environment are depicted in Fig.~\ref{fig_model1}.

The agent is assumed to start at the origin and walk in a straight line along a given radial direction $\theta_{\mathrm{0}}$. At each time step $i$ the agent moves in a random direction $\theta_i$ (see Fig.~\ref{fig_model1}A), relative to its current heading, i.e. $\theta_i=\theta_{i-1}+\delta \theta$ where for simplicity $\delta \theta$ is drawn from a uniform distribution $[0, \theta^*]$ \cite{Codling:2008bm} (choosing from a Gaussian distribution does not change any of results qualitatively). The agent stops at certain intervals to reorient, i.e. it resets $\theta_i=\theta_{\mathrm{0}}$, which we assume takes time, given that the agent has to take a bearing (corresponding to the beetle reorienting on the dung ball). Then, after $N$ steps that include $N_r$ reorientations, the end--to--end vector of the agent is $\textbf{R} = \sum_{i=1}^N \textbf{dr}_{i}$ where $\textbf{dr}_{i}=(\sin\theta_i, \cos\theta_i)$. 

To characterize the competition between accuracy and speed associated with the task of moving as far from the origin as quickly as possible, we define a simple cost function that penalizes the deviation from a straight line and also penalizes the time spent for reorientation  
\begin{equation}\label{eq:cost}
f=\frac{N-\left|\textbf{R}\right|}{N} + \frac{N_{r}}{N}.
\end{equation} 
Frequent reorientations result in $\left| {\textbf{R}} \right| \rightarrow N $ and leads to the first term being small, but the second being large. In contrast, few reorientations will make the second term small, while the first will be large. Continuity suggests that the cost will be a minimum when the the number of steps before a reorientation event $n=N/N_r$, or equivalently, the frequency of reorientation, is an optimum. Before proceeding to find this, we point out that this problem has some similarities to a recent class of problems named ``search with reset''' \cite{Evans:2011jo,Evans:2011el}, but differs qualitatively in that here we consider diffusion in orientational rather than rectilinear space, while coupling reorientation to translational motion, making an analytic approach difficult except in the simplest of situations.   

We start with the simplest variant of the question of optimal switching, assuming that the heading direction is reset to zero every $n$, with  $N=nN_r$, {complete attention to navigational cues between reorientation events}, i.e. $\tau=1$, so that there is no error amplification, i.e. $A=1$, and finally there is no acquisition error, i.e. $\epsilon_d=0$. Later we will include the additional stochasticity associated with randomness in the choice of reorientation intervals, and errors in acquisition, {variable} attention span, and accompanying error amplification.  

\begin{figure}
   \centering
   \includegraphics[width=12cm]{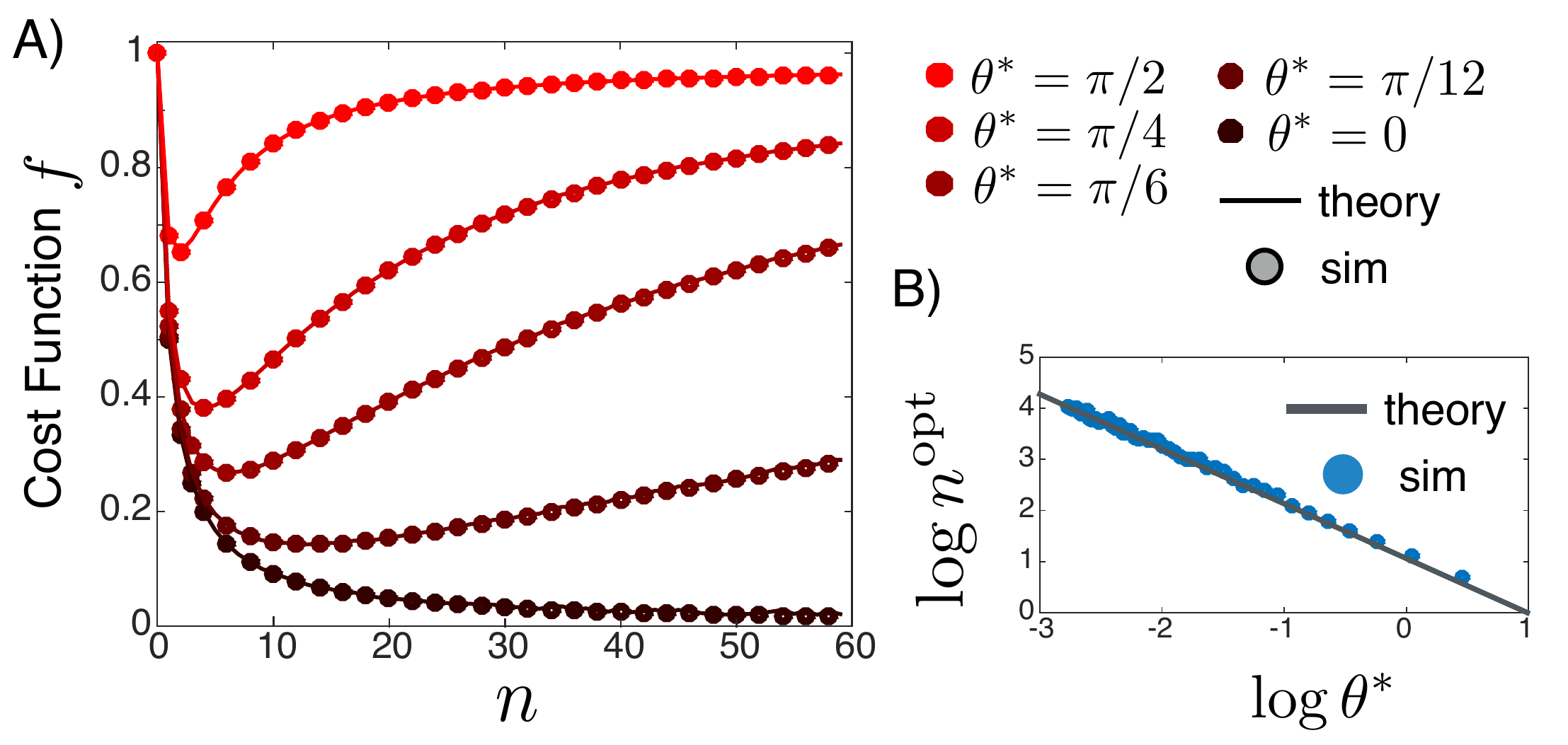} 
   \caption{Cost function vs.~$n$, for simulations with $5000$ different realizations, $N=1000$. The curves are colored according to their $\theta^*$ value (see legend). Theoretical prediction shown as solid lines, and numerical values obtained via simulations shown as open circles. B) Loglog plot of $n^{opt}(\theta^*)$ vs.~$\theta^*$ derived from both the theoretical prediction (red solid line), and simulations (open circles).} 
  \vspace{-10pt}  \label{fig_cost_sim_and_th}
\end{figure} 

With these assumptions, the agent's path is a correlated random walk (CRW) in between reorientation events, and a biased random walk at the long time scale (BRW) \cite{Codling:2008bm}, as illustrated in Fig.~\ref{fig_model2}A--B. The path always starts with a step along $\theta_0$, followed by $n-1$ steps CRW, resulting in a mean square end--to--end distance $\left<R^2_{CRW}\right>$ and mean deviation angle $\theta^{CRW}$. These two quantities define a sector accessible to the agent between reorientation events with central angle $2\theta^{CRW}$ and radius $\sqrt{\left<R^2_{CRW}\right>}$ (Fig.~\ref{fig_model2}C). Then, following \cite{Cheung:2007hna}: 
\begin{equation}\label{eq:RCRW2}
\left<R^2_{CRW}\right>= n -1 + \frac{2\beta}{1-\beta}\left(n -1 - \frac{1-\beta^{n-1}}{1-\beta}\right)\end{equation} \begin{equation}\label{eq:RCRWx} \left<R_{\parallel CRW}\right>=\frac{\beta-\beta^{n}}{1-\beta}\end{equation} \begin{equation}\label{eq:thetaCRW}
\theta_{CRW} = \cos^{-1}\left(\frac{\left<R_{\parallel CRW}\right>} {\sqrt{\left< R_{CRW}^{2}  \right>}} \right)
\end{equation} where $n$ is the number of steps before a reorientation, and $\beta=\left<\cos\theta_i\right>$, $\left<R_{\parallel CRW}\right>$ is the mean distance traveled along the directional bias, and $\theta_{CRW}$ is the mean deviation from the intended direction.

On longer time scale, the entire random walk may be treated as a BRW with individual steps given by the CRWs (Fig.~\ref{fig_model2}C). The end--to--end distance of a BRW is defined as $\left<R^2_{BRW}\right> = \left<R^2_{BRW\parallel}\right> + \left<R^2_{BRW\perp}\right>$, where $\left<R^2_{BRW\parallel}\right>$ and $\left<R^2_{BRW\perp}\right>$ are the mean square displacements parallel and perpendicular to the directional bias, respectively. Following \cite{Cheung:2007hna}: 
\begin{equation} \left<R^2_{BRW\perp}\right> = N_r\left(1- \gamma_{BRW} \right) \left<dr^2_{BRW}\right>
\end{equation} 
\begin{equation} \left<R^2_{BRW\parallel}\right> = N_r^2 \gamma_{BRW} \left<dr^2_{BRW}\right> + N_r^2
\end{equation} 
where $\gamma_{BRW}=\left<\cos^2\theta_{CRW}\right>$ and the mean square step size $\left<dr^2_{BRW}\right>=\left<R^2_{CRW}\right>$. Here, we have added the $N_r$ term to $\left<R^2_{BRW\parallel}\right>$ to account for the fact the first step after reorientation is parallel to the directional bias. Therefore, the mean square displacement of the entire walk is: \begin{eqnarray}
\left<R^2_{BRW}\right> &&= N_r\left(1- \gamma_{BRW} \right) \left<R^2_{CRW}\right> \nonumber\\ && + N_r^2 \gamma_{BRW}\left<R^2_{CRW}\right> + N_r^2
\end{eqnarray}

Using ~(\ref{eq:RCRW2}) and ~(\ref{eq:thetaCRW}) in the above relation allows us to calculate $|{\bf R}|=\left<R^2_{BRW}\right>^{1/2}$, so that the cost given in ~(\ref{eq:cost}) becomes: 
\begin{eqnarray}
f\left(N,n,\beta(\theta^*) \right) = \frac{N-\sqrt{\left<R^2_{BRW}\right>}} {N} + \frac{1}{n} 
\label{eq:wideeq}
\end{eqnarray}\begin{eqnarray}
= 1 + \frac{1}{n} -\frac{\sqrt{N}}{N(\beta -1) n} \sqrt{N \left(\beta ^n-2 \beta +1\right)^2   -\left(\beta ^2-1\right) n^2-n \left(\beta ^n-1\right) \left( \beta ^n-2 \beta-1\right)\nonumber} 
\label{eq:wideeq}
\end{eqnarray}

In Fig.~\ref{fig_cost_sim_and_th}, we show the dependence of the cost given by (\ref{eq:wideeq}) on the frequency of reorientation $n$. When the width of the turning angle distribution $\theta^*=0$, the agent travels along a perfectly straight line and the cost function $f$ decreases monotonically with $n$ (black solid curve). For a uniform turning angle distribution, $\beta={\sin\theta^*}/{\theta^*}$, so that when $\theta^*\neq0$, $f$ is optimal for a particular $n(\theta^*$ (colored solid curves). We confirm these analytical results using simulations, as shown in Fig.~\ref{fig_cost_sim_and_th}A. 

To study how the optimal reorientation frequency, $n^{opt}$ depends on $\theta^*$, we consider the cost function in the asymptotic limit $f\left(N \to \infty,n,\beta(\theta^*) \right)$ 
\begin{equation} 
\lim_{N \to \infty} f= \frac{\left(\beta-1\right) \left(1 + n\right) + 1 - \beta^n}{(\beta-1) n}
\end{equation} 
Minimizing this with respect to $n$, yields \begin{equation} \label{eq:opt_n}
 n^{opt}=\frac{1 + W(-\beta/e)}{-\log\beta} 
 \end{equation} where $W$ is the Lambert W--function (the solution of $z=W\exp{W}$). For the case when the turning angle distribution is uniform: \begin{equation} 
 n^{opt}(\theta^*)=\frac{1 + W(-\frac{\sin\theta^*}{\theta^*e})}{-\log(\sin\theta^*/\theta^*)}  \sim \frac{\pi}{\theta^*}
 \end{equation} 
 (see Appendix for a derivation). In Fig.~\ref{fig_cost_sim_and_th}B we plot $n^{opt}(\theta^*)$ vs.~$\theta^*$ derived from our simulations, and see that they agree well with our simple theoretical prediction. Furthermore, Fig.~\ref{fig_cost_sim_and_th}A shows that as $\theta^*$ increases it becomes more crucial to be close to the optimal reorientation interval, as small deviations in $n$ result in a sharp increase of the cost.  
 
\begin{figure}
   \centering
   \includegraphics[width=12cm]{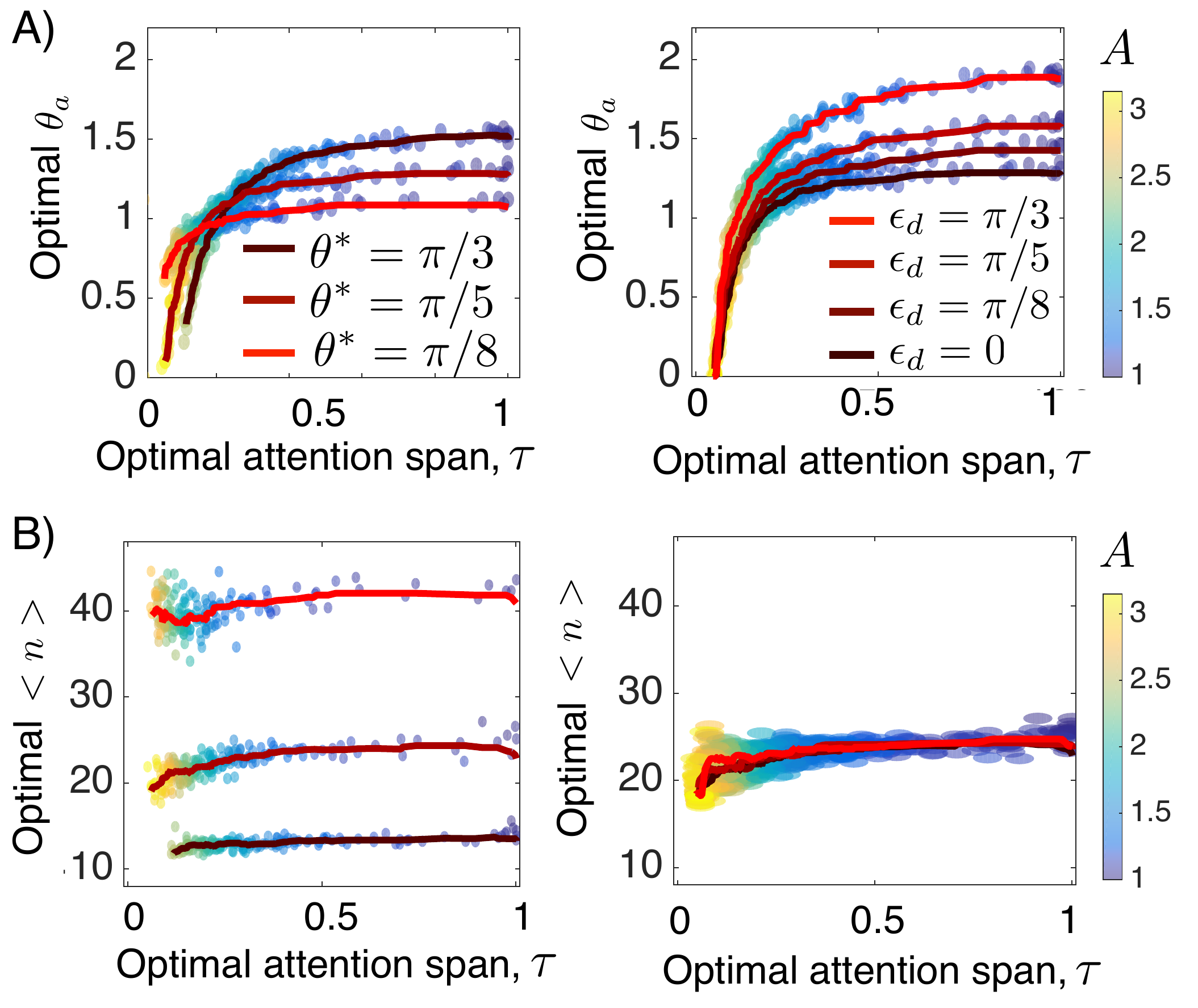} 
   \caption{{Optimal strategies in stochastic navigation. (A) Optimal activation angle $\theta_a$ and attention span $\tau$.  B) The optimal mean reorientation interval, $<n>$, vs.~$\tau$. Left and right panels, summarize the effect of $\theta^*$, and $\epsilon_d$, respectively on $\theta_a, \tau$. Each point in the scatter plots corresponds to a set of parameters controlled by the agent or the environment, as summarized in Fig.~1. Color scale corresponds to the attention error amplification, $A$. Solid lines created using a Gaussian filter over the scatter points as shown in legend. $N=100$; $\lambda_{CMA} = 10$ (population size per generation).}} 
  \vspace{-10pt}  \label{fig_cma}
\end{figure}

Having understood this simple scenario, we now go back to address the complexities associated with noise in acquisition, planning and execution, in addition to the noise in the turning angle distribution $\theta^*$. To address this, we use numerical simulations with the acquisition error $\epsilon_d \neq 0$, the error amplification associated with {attention} $A>1$, and ask how the agent must optimize the attention span $\tau \neq 1$ and the threshold activation angle for reorientation $\theta_a$ to minimize the cost function defined in Eq.~\ref{eq:cost}. We use the covariance matrix adaptation algorithm (CMA--ES) \cite{Hansen:2006gz} to determine the optimal strategy, where given a set $(\theta^*, A, \epsilon_d)$, we determine the { optimal set $(\theta_a,\tau)$.} The CMA-ES is a stochastic derivative--free optimization method for non--linear or non--convex continuous optimization problems. By incrementally increasing the probability of previously successful candidate solutions, we iteratively perform the following three steps: (i) sample p new sets of $(\theta_a,\tau)$ following the distribution with the updated mean and covariance of the cost function $f$ (with step size built in), (ii) evaluate $f$ and re--order the sampled solutions based on their fitness, (iii) update the internal state variables $(\theta_a,\tau)$, including the mean, the isotropic and anisotropic evolution path, the covariance matrix, and the step size, based on the q best out of p solutions, until we have converged to the minimal cost that will yield $\tau=\tau(\theta^{*},A, \epsilon_d)$, $\theta_a=\theta_a(\theta^{*},A, \epsilon_d)$.   

The results of our optimization are summarized in Fig.~\ref{fig_cma}. We see that both the optimal attention span $\tau$ and optimal activation angle $\theta_a$ decrease monotonically with increase in the { attention} error amplification $A$ consistent with our intuition. { However, the mean reorientation interval $<n>$ (extracted retrospectively from the simulations), is relatively independent of $A$ (Fig.~\ref{fig_cma}B)}. When the variance of the turning angles $\theta^*$ increases, we see two regimes; for large $A$, $\tau$ is large and $\theta_a$ increases with $\theta^*$. For small values of $A$ there is no clear trend. Finally, $\theta_a$ and $\tau$ increase monotonically with $\epsilon_d$.  




Inspired by the switching strategy of the dung beetle, we have posed and solved an optimization problem of geocentric navigation interspersed by egocentric cue integration. In the simplest setting, we find that the optimal reorientation interval is inversely proportional to the environmental noise and is invariant to sensory acquisition noise.  In more complex settings, our study highlights the variations in the optimal navigation strategy that balances accuracy, speed and effort. 
 Finally, it is worth noting that our results may be applicable to a more broad range of stochastic navigational problems are combined with constraints. A natural next step would be to test the theory with animals (e.g. dung beetles) as well as autonomous robotic agents.  

\begin{acknowledgments}
We thank the MacArthur Foundation for partial financial support, { Elisabetta Matsumoto for a discussion on the asymptotics of the Lambert W function, and the Mahadevan lab for comments.}
\end{acknowledgments}

\bibliography{Xbib}

\appendix\section{Manipulation of Lambert W-function}

Here we show that
\begin{equation}
\lim_{\theta^*\rightarrow0} \frac{1+W\left(-\frac{\sin(\theta^*)}{e \theta^*}\right)}{-\log\left(\sin(\theta^*)/\theta^*\right)}=\frac{2\sqrt{3}}{\theta^*}+O(\theta^*),
\end{equation}
where $W(z)$ is the Lambert W-function.
First we note a few identities of the $W(z)$:
\begin{eqnarray}
\frac{d W(z)}{dz} &=& \frac{W(z)}{z(W(z)+1)}\\
\lim _{z\rightarrow-\frac{1}{e}} W(z)&\sim& -1+\sqrt{2 e}\sqrt{z+\frac{1}{e}}+O(z).
\end{eqnarray}


Since $\lim_{\theta\rightarrow0}\sin(\theta)/\theta\sim1-\frac{6}{\theta^2}$, an expansion near $W(z\rightarrow-\frac{1}{e})$ is appropriate.  Therefore, the numerator is:
\begin{eqnarray}
1+W\left(-\frac{\sin(\theta^*)}{e \theta^*}\right)&&=1-1+\sqrt{2e}\sqrt{\frac{1}{e}{\frac{{\theta^*}^2}{6}}}\\ \nonumber&&+O\left({({\theta^*})^2}\right)\\ \nonumber&&=\frac{\theta^*}{\sqrt{3}}+O\left({({\theta^*})^2}\right).
\end{eqnarray}
Likewise, the expansion for the denominator is, $\log(\frac{\sin(\theta^*)}{\theta^*})=-\frac{6}{{\theta^*}^2}+O\left({({\theta^*})^0}\right).$
Combining these, we demonstrate
\begin{equation}
\lim_{\theta^*\rightarrow0} \frac{1+W\left(-\frac{\sin(\theta^*)}{e \theta^*}\right)}{-\log\left(\sin(\theta^*)/\theta^*\right)}=\frac{2\sqrt{3}}{\theta^*}+O(\theta^*).
\end{equation}

\end{document}